\documentclass[fleqn,twoside]{article}
\usepackage{espcrc2}
\usepackage{epsfig,graphics}

\newcommand{\chiPT}{$\chi{\rm PT}$}
\newcommand{\Tr}{\mbox{Tr}}

\def\psibar{{\overline\psi}}
\def\order{{\cal O}}

\title{Topology and Low Lying Fermion Modes}

\author{Robert G. Edwards\address[JLab]{Jefferson Lab\\ 
        12000 Jefferson Avenue\\ 
        MS 12H2, Newport News, Virginia, USA}}
       
\begin{document}

\begin{abstract}
Recent results concerning the relation of topology and low-lying fermion
modes are summarized.

\end{abstract}

\maketitle

\section{Introduction}

Low lying fermionic modes are believed to play an important r\^ole in
QCD. This review covers some of the recent
developments in the study of the relationship of low lying fermion
modes with topology in QCD. The study of topology in general is not
covered here. In particular, fermionic eigenmodes (including
zero modes) are important in 3 areas discussed here: (1) low lying
modes in particle spectrum, (2) low lying modes in global topology
(e.g., chiral fermions), (3) low lying modes in the implementation  
of chiral fermions.
The use of fermionic modes to probe for possible mechanism of chiral
symmetry breaking in QCD are possible. The recent advent of the
construction of chiral fermions on the lattice are crucial in new
studies.

\section{Spectrum}

\subsection{Eigenmodes in Spectrum}
One expects low lying fermionic eigenmodes to be important in quark
propagation because they should dominate the spectral sum for small
enough quark masses. By computing some number of the lowest lying
eigenvectors of the Dirac operator one can expect that most of the
physics of the quark propagator can be obtained. To make these
statements more precise, an example is given of the computation of the
$\eta'$ mass. 
Consider the spectral decomposition of a quark
propagator -- for simplicity use a hermitian form of the Dirac
operator
\begin{equation}
H=\gamma_5 D_w, \quad H\psi_i(x) = \lambda_i \psi_i(x)
\end{equation}
where $D_w$ is the standard Wilson-Dirac operator, $H$ is the
hermitian Wilson-Dirac operator and the $\psi_i(x)$ are the
eigenvectors with eigenvalue $\lambda_i$. Then the spectral
representation of the quark propagator is
\begin{equation}
H^{-1}(x,y) = \sum_i \frac{\psi_i(x) \psi^\dagger_i(y)}{\lambda_i}\quad .
\end{equation}

\vskip -11cm
\rightline{JLAB-THY-01-33}
\vskip +10.2cm

The correlation function for the $\eta'$ involves a disconnected piece
which is typically stochastically estimated. A possible improvement is
to truncate the spectral sum with the lowest few eigenvectors (which
should give the largest contribution) and stochastically estimate the
remainder. The idea is to represent $H = \sum_i H_i + H_\perp$ where
$H_\perp$ are the remaining modes. Then in all relations where
$H^{-1}$ enters, find all the terms involving only $\sum_i H_i$ and
the cross terms involving $H_\perp$. Since the eigenvectors contain
the full information (in the subspace) of the propagator (not just
emanating from a single point), impose translation symmetry by summing
across the lattice for the lowest modes. This sum gives a volume times
more statistics for the modes (hopefully) dominating the correlation
function. For the remaining terms involving $H^{-1}_\perp$ use a
stochastic estimator. The linear system solutions involved in this
latter step are accelerated since the condition number is lowered from
projecting out the lowest eigenmodes.

How many eigenmodes are needed? The scale should be set by the chiral
condensate, hence the number should grow at least like the volume.

Instead of computing the lowest eigenmodes, deflation can be used to
compute the eigenvectors at the time of the linear system solution
thus accelerating the stochastic estimation~\cite{Wilcox_01}.

\begin{figure}[t]
\epsfig{file=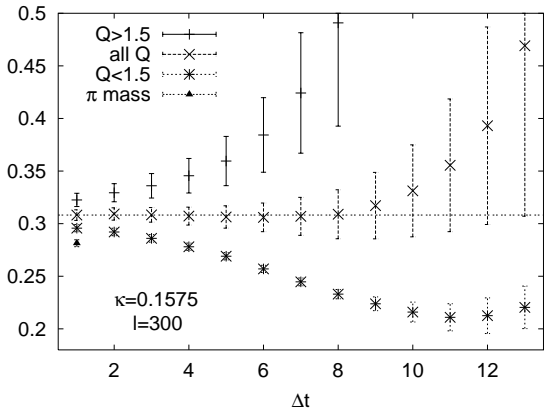, clip=, width=8cm}
\vskip -8mm
\caption{Dependence of $\eta'$ effective mass on topological charge
(Ref.\cite{Wupp_01}).}
\label{fig:etamass}
\end{figure}

\begin{figure}[t]
\epsfig{file=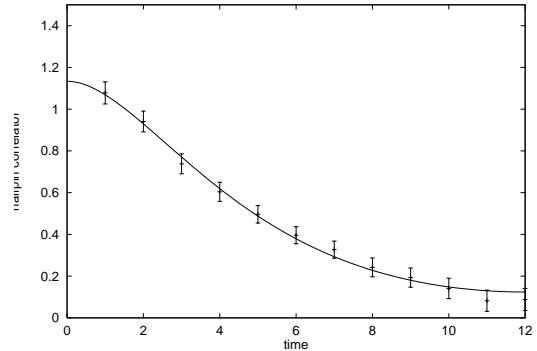, clip=, angle=270, width=7cm}
\vskip -7mm
\caption{One-parameter fit to a pure double-Goldstone pole form for
the $\eta'$ hairpin correlator. The pion mass is fixed from the
valence propagator analysis 
(Ref.\cite{FNAL_00}).}
\label{fig:hairpin_fit}
\end{figure}

One obvious question is whether to use for Wilson-like fermions the
hermitian or non-hermitian operator. A comparison of the truncated
spectral sum of the pion propagator shows poor convergence in the
non-hermitian version when adding more eigenvectors~\cite{Wupp_01}.
The hermitian version shows slow but even convergence.  For chiral
fermions, the choice is irrelevant.

The Wuppertal group has used the truncated spectral version of the
hermitian quark propagator for a calculation of the $\eta'$
correlation function using $N_f=2$ dynamical Wilson
fermions~\cite{Wupp_01}.  A similar calculation was made for staggered
fermions~\cite{Kilcup_97}.  One goal is to determine the topological
charge dependence of the $\eta'$ mass.  Using a fermionic definition
of the charge and binning configurations according to $Q$, the
effective masses show a $Q$ dependence as seen in
Fig.~\ref{fig:etamass}. They find the truncated spectral sum method
competitive with stochastic estimation for their large quark
masses. The spectral sum method will not suffer dramatically going to
smaller quark masses, however. The truncation of the spectral sum is
evident in the not very cosh-like behavior of the $\eta'$ correlation
function. The calculation could be made exact by including the extra
stochastic estimation.

While not using a fermionic method, a calculation by the UKQCD
collaboration of the flavor singlet mesons with $N_f=2$ dynamical
$\order(a)$ improved Wilson fermions has been made~\cite{UKQCD_ozi}. While
statistically noisy, the OZI rule of suppressed singlet -- non-singlet
mass splittings is seen like in the quenched case~\cite{Thacker_00}.

\section{Topology}

\subsection{Topological Susceptibility}

The topological susceptibility is intimately tied with low lying
fermion modes and is an important probe of the QCD vacuum relating to
the $U_A(1)$ problem. Pure gauge calculations of the topological
susceptibility are consistent with the Witten-Veneziano
prediction~\cite{Giusti_01}. Another important test is to determine
how the susceptibility depends on the quark mass.  There are several
recent large calculations of the topological susceptibility using
$N_f=2$ dynamical fermions, from CPPACS (mean-field
clover)~\cite{sus1}, UKQCD (non-pertubative clover)~\cite{sus2},
SESAM/T$\chi$L (Wilson)~\cite{sus3}, and thin-linked
staggered~\cite{sus4}. By continuum arguments, the susceptibility is
expected to behave like
\begin{equation}
\chi(m) = \frac{\Sigma m}{N_f}(1 + \order(m)) 
 = \frac{m_\pi^2 F_\pi^2}{2 N_f}(1 + \order(m)).
\end{equation}

\begin{figure}[t]
\begin{center}
\epsfig{figure=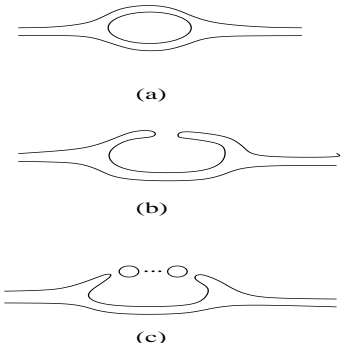, clip=, width=4.5cm, height=4.5cm} 
\end{center}
\vskip -8mm
\caption{Contributions to the $a_0$ propagator
from an $\eta-\pi$ intermediate state.}
\vskip -2mm
\label{fig:quark_lines}
\end{figure}

Only the UKQCD calculation claims to unambiguously see the expected
decrease of $\chi(m)$ by lowering $m$. A comparison of the various
results at finite lattice spacing requires taking into account
discretization errors~\cite{Durr_01}. Na\"ive linear extrapolations in
$m_\pi^2$ (fixing $F_\pi$) give poor fits, suggesting that
discretization effects are large.  Besides large quark masses used
there are other sources of systematic errors including the definition
of $\chi(m)$ which involves contact terms and mixing with the unit
operator~\cite{Giusti_01}. For a chiral fermion action with a
fermionic definition of $\chi(m)$ these counter-terms are absent only
in the $N_f=2$ chiral limit. They could be avoided by using a
finite-volume scaling technique~\cite{DEHN_99} for small $m$ applied
to $\chi(m)$.

\begin{figure}[t]
\epsfig{figure=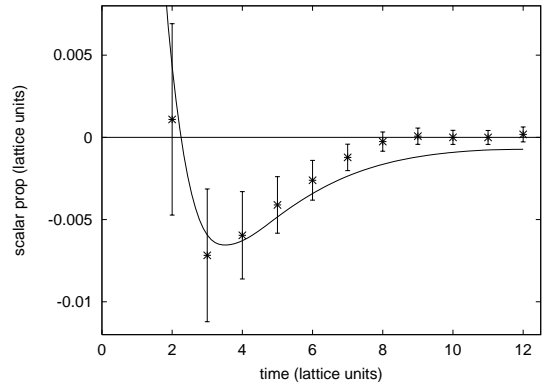, clip=, angle=270, width=7.5cm}
\vskip -6mm
\caption{Comparison of scalar $a_0$ propagator with
the bubble sum formula 
fitted to the interval t=1-6
(Ref.\cite{FNAL_01}).}
\label{fig:bubble_fit}
\end{figure}

\subsection{Quenched Pathologies}

Another area low lying eigenmodes play an important role are in
quenched pathologies.  How well is QCD described by an effective
chiral theory of interacting particles (e.g., pions in chiral
dynamics)? Suppressing the fermion determinant leads to well known
pathologies as studied in chiral perturbation theory~\cite{chipt}.
The pathologies are manifested in the $\eta'$ propagator
missing vacuum contributions. A new dimensionful parameter is induced
in \chiPT.  Power counting rules are changed leading to new chiral
logs and powers terms.  The predicted pathologies for the pion mass
were studied with Wilson fermions by
CPPACS~\cite{CPPACS_99}. There are recent studies with Wilson fermions
in the Modified Quenched Approximation \cite{FNAL_00,FNAL_01} and
using the overlap chiral fermion operator~\cite{KY_01}.

The pion mass was one of the first observables studied for the effects
of quenched chiral logs; however, it is not necessarily the easiest place
to look. The additional quenched logs appear in the pion mass from the
missing loop contributions in the $\eta'$ correlator. In
\cite{FNAL_00}, the unique piece of the $\eta'$ correlator --
the hairpin -- was computed directly and compared to quenched
\chiPT. Fig.~\ref{fig:hairpin_fit} shows the correlator along with the
fit holding $m_\pi$ fixed to the
simple mass insertion formula given by
\begin{eqnarray}
&&\!\!\!\!\!\!\!\!\!\!\!\!\int \frac{d^4 p}{(2\pi)^4} \; e^{i p\cdot x}\,
\langle \Tr\gamma_5 G(x,x)\Tr\gamma_5 G(0,0)\rangle = \cr
&&\!\!\!\!\!\!\!\!\!\!\!\!\qquad f_P\frac{1}{p^2+m^2_\pi} m_0^2\frac{1}{p^2+m^2_\pi} f_p
\end{eqnarray}
The correlator is well described by this form with the only free
parameter the $\eta'$ mass insertion $m_0^2$. The pseudoscalar decay
constant also shows the expected quenched artifact given by
\begin{equation}
f_p^{quenched} = \left( \frac{1}{m^2_\pi}\right)^\delta {\tilde f}_P
\end{equation}
while the axial vector decay constant does not show such a divergence.
After considering various observables, the overall determination of
$\delta=0.065$(13) compared to $\delta=0.18$ in Ref.\cite{chipt}.

In Ref.~\cite{KY_01}, the overlap-Dirac operator was computed on
a $20^4$ lattice at a lattice spacing of $a=0.13$fm. Quenched chiral
logs were observed in $m_\pi^2$ and $f_P$ giving $\delta\sim 0.15$ to $0.4$.

A dramatic example of quenching is exhibited in the non-singlet scalar
particle~\cite{FNAL_01}. For the $a_0$, there is an $a_0 \rightarrow
\eta-\pi$ intermediate state as seen in
Fig.~\ref{fig:quark_lines}. The vacuum contribution (a) is missing in
the quenched approximation along with the vacuum bubbles on the
hairpin piece (c), but a remaining hairpin piece (b) is present.  The
remaining hairpin piece has a quadratic divergence coming from the
three pion propagators present. This quenched artifact causes a loss
of positivity of the propagator as seen in
Fig.~\ref{fig:bubble_fit}. The fit is to the 1-loop bubble term as
computed in quenched \chiPT\ and is in good agreement with the
data~\cite{FNAL_01}.  The difficulty in this Wilson fermion (MQA)
approach is the mixing of would-be zero modes and non-zero modes and
in particular what are the contributions from each part. For example,
is the lack of positivity a zero mode effect? It is not expected to
be, but more will be discussed on zero modes in the next section.

While the $a_0$ quenched artifact can be viewed as the remnants of a
decay, a true decay requires full QCD. The MILC collaboration has used
an improved staggered fermion action in a $N_f=2+1$ calculation of the
$a_0$ decay~\cite{MILC_01}.  
Evidence is found for a level splitting in the $a_0$ mass and an $\eta+\pi$ 
state.

\subsection{Condensate}

Several model calculations indicate the quenched chiral condensate
diverges at $T=0$ (e.g., Ref.~\cite{cond}).
Damgaard~\cite{Damgaard_01}, shows via quenched $\chi$PT that the first finite
volume correction to the chiral condensate diverges logarithmically in
the 4-volume.  Some simple relations for susceptibilities of
pseudoscalar and scalar fields can be derived in the continuum and on
the lattice with a chiral fermion action (see for example Ref.\cite{EHN_98}). 
Define $\pi$ and $a_0$ fields by
\begin{equation}
\pi^a(x) = i\psibar(x)\gamma_5 \tau^a\psi(x), 
a_0(x) = -\psibar(x)\tau^a\psi(x).
\end{equation}
Then the $\pi$ and $a_0$ susceptibilities are given by
\begin{eqnarray}
\sum_x\langle\pi^a(x)\pi^a(0)\rangle &=& \frac{1}{m}\langle\psibar\psi\rangle\cr
\sum_x\langle a_0^a(x) a_0^a(0)\rangle &=& 
\frac{1}{m}\langle\frac{d}{dm}\langle\psibar\psi\rangle_A\rangle\quad .
\end{eqnarray}
Note in the quenched approximation the $a_0$ susceptibility is given by
the derivative of the chiral condensate. These
relations are true including and excluding global topology terms.

The Banks-Casher relation on a finite lattice is
\begin{equation}
\frac{1}{V}\sum_x\langle\psibar(x)\psi(x)\rangle 
 = \frac{|Q|}{mV} + \frac{1}{V}\sum_n f(\lambda_n,m),
\end{equation}
where $f$ is some unspecified function of the non-zero eigenvalues.
One sees the global topology term is irrelevant in the thermodynamic limit.
Taking the appropriate limits the usual relation is recovered
\begin{equation}
\lim_{m\rightarrow 0}\lim_{V\rightarrow\infty}
\frac{1}{V}\sum_x\langle\psibar(x)\psi(x)\rangle = \pi\rho(0^+)\quad.
\end{equation}

If the chiral condensate diverges without the topology contribution,
the $a_0$ susceptibility must be negative and diverge. This divergence
requires large enough physical volume to be apparent. Difficulties in
the past include: (1) mixing of (would-be) zero and non-zero modes
with staggered fermions, (2) mixing of topology and non-zero modes and
contact terms with Wilson fermions, (3) until recently chiral fermion
studies not on large enough lattices, e.g., random matrix model tests,
spectrum tests, direct measurement tests.  More direct tests of the
divergence of the condensate will come from thermodynamics.

\begin{figure}[t]
\epsfig{figure=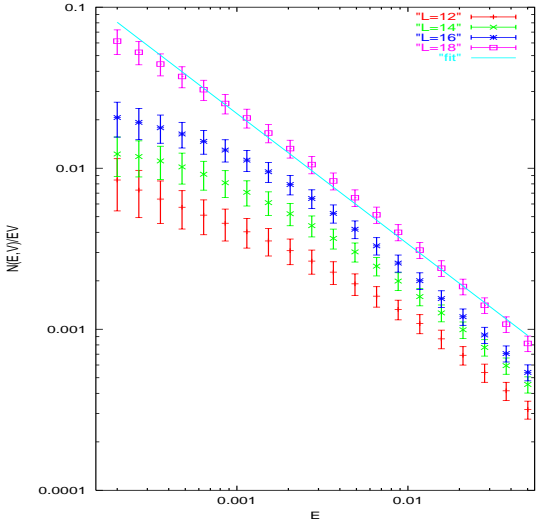, clip=, width=7.5cm, height=7cm}
\vskip -5mm
\caption{A plot of $N(E,V)/(EV)$ from the
small nonzero eigenvalues on four different lattices
plotted on a log-log scale. The solid line is a
least squares fit to the $L=18$ data weighted by
the statistical errors (Ref.\cite{KN_01}).}
\label{fig:diverging_chi}
\end{figure}

\subsection{Thermodynamics}

The deconfined phase of SU(2) quenched gauge theory, $L^3\times 4$,
$\beta=2.4$, above $N_t=4$ transition was studied using the
overlap-Dirac operator~\cite{KN_01}. A study of the build-up of
the density of eigenvalues near zero, $\rho(E)$ indicates the 
quenched chiral condensate (without topology) is
diverging~\cite{KN_01}. Define the density of zero eigenvalues
$\rho(E)$ from the derivative of the cumulative distribution
\begin{eqnarray}
N(E,V) &=& \#(\lambda > 0)\quad {\rm where\ }\lambda < E \cr
\rho(E) &=& \frac{d}{dE}\lim_{V\rightarrow\infty} \frac{N(E,V)}{V}\quad .
\end{eqnarray}
As seen in Fig.~\ref{fig:diverging_chi}, the cumulative distribution appears to
continually rise and track line on a log plot -- hence the derivative
(the condensate) diverges with increasing lattice size.

A diverging condensate indicates the spectral gap is closed. However,
a decrease in the topological susceptibility is seen when crossing to
$T > 0$.  Has the quenched approximation distorted the vacuum so much
that it has invalidated continuum arguments as to the nature of the
$T>0$ phase transition or mechanisms of confinement? The claim is 
no -- there is merely a large accumulation of near zero fermion
modes. However, models predict changes in the vacuum structure
crossing to the deconfined and (supposedly) chirally restored phase
and low lying fermion modes can again be used to probe the vacuum.

\subsection{QCD Vacuum}

It is generally accepted that the QCD vacuum is characterized by
strongly fluctuating gluon fields with clustered or lumpy distribution
of topological charge and action density.  Confinement mechanisms are
typically ascribed to a dual-Meissner effect -- condensation of
singular gauge configurations such as monopoles or vortices (see
review Ref.~\cite{Perez_00}). Disagreements of various
models include:
(1) instanton models provide chiral symmetry breaking, but not confinement,
(2) center vortices provide confinement and chiral symmetry breaking,
(3) composite nature of instantons (linked by monopoles - calorons) at $T>0$.
A different approach relying on properties of gauge field correlators and
not specific gauge field types describes a large class of non-perturbative 
phenomena~\cite{Simonov_00}.

Singular gauge fields are probably intrinsic to SU(3) (e.g., not just
Gribov copies associated with gauge fixing). These singularities 
impose boundary conditions on quark and gluon fluctuations which moderates the
QCD action. 
For example, instantons have locked chromo-electric and magnetic fields $E^a
= \pm B^a$ that decrease in strength in a certain way. 

In a hot configuration one expects huge contributions to the action beyond such
special type of field configurations. However, there could possibly be 
domains of (near) field locking. In recent calculations~\cite{Kall_01}, these
domains have been shown to be sufficient to produce chiral symmetry
breaking, and confinement (area law).

\subsection{Instanton Dominance in QCD?}

Witten has argued~\cite{Witten_79} that topological charge
fluctuations are clearly involved in solving the $U_A(1)$
problem. However, the dynamics of the $\eta'$ mass generation need not
be associated with semiclassical tunneling events. In particular,
large vacuum fluctuations from confinement also produce topological
fluctuations.  Large $N_c$ arguments are incompatible with instanton
based phenomology - namely instantons produce and $\eta'$ mass that
vanishes exponentially, but from large $N_c$ chiral dynamics suggest
that $m^2_{\eta'}\sim 1/ N_c$.  Witten speculated that the $\eta'$
mass comes from the coupling of $U_A(1)$ anomaly to topological charge
fluctuations and not instantons.

\begin{figure}[t]
\epsfig{figure=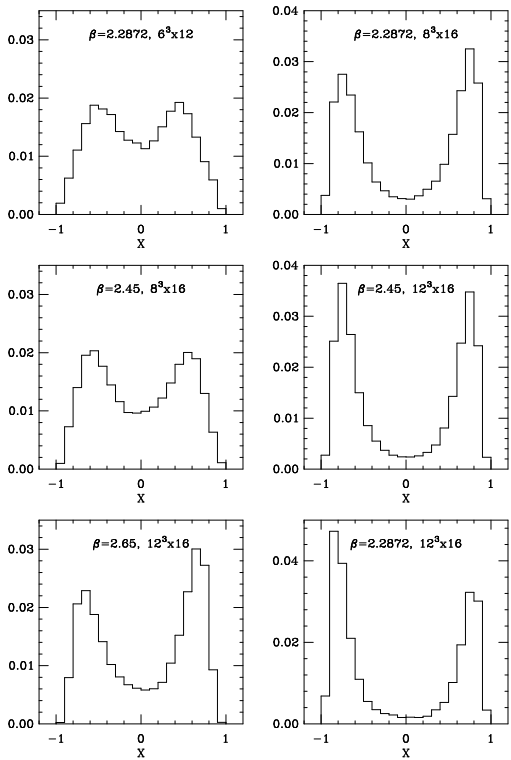, clip=, width=7cm}
\vskip -10mm
\caption{Chirality histograms for the lowest two non-zero modes of the
overlap-Dirac operator at the 2.5\% sites with the largest
$\psi^\dagger \psi(x)$ on the six ensembles with Iwasaki gauge action.
The systems in the left column all have approximately the same volume in
physical units. The systems in the top two panels in the right column also
have the same, about a factor of 3 larger, volume, while the system in the
lower right hand corner panel has the largest volume (Ref.~\cite{EH_01}).}
\label{fig:hist_IW_2l_2p5pc}
\vskip -9mm
\end{figure}

\subsection{Local Chirality}

A test of Witten's conjecture was put forward with a 
local measure of chirality of non-zero modes~\cite{Horvath_01}. 
The relative orientation of the left and right handed
components of low lying eigenmodes is used
\begin{equation}
\tan\left(\frac{\pi}{4}(1+X(x))\right) 
 = \sqrt{\frac{\psi^\dagger_L(x)\psi_L(x)}{\psi^\dagger_R(x)\psi_R(x)}}\quad .
\end{equation}
Here $X(x)$ is a site dependent measure of the local chirality. A
completely chiral state (an exact zero mode) have for some sites
$X(x)=\pm 1$.  These regions of local chirality come from (near)
locking of the chromo-electic and magnetic ($E$ and $B$) fields.
Wilson fermions were used as a test~\cite{Horvath_01}, and since
the chirality appeared random the claim was there is no instanton
dominance in the QCD vacuum.  A flurry of papers using improved
improved actions followed~\cite{Flurry,EH_01}.

The chirality histogram for the overlap-Dirac operator is shown in
Fig.~\ref{fig:hist_IW_2l_2p5pc}.  Shown is the histogram of $X(x)$ for
2.5\% of the sites with the largest $\psibar\psi$ at three physical
volumes keeping only the lowest two non-zero modes. Keeping more small
non-zero modes indicates a finite density of such chiral peaked modes
which may survive the continuum limit~\cite{EH_01}. 
The mixing (trough) in the
histograms are not related to dislocations. Namely, comparing
histograms from different gauge actions and vastly different gauge
field dislocation content appear identical indicating mixing comes
from large sized eigenmodes. In addition, no significant peaking was
found in a study of $U(1)$ in the confined phase where instantons are
not present~\cite{Berg_01}.

These results are all consistent with instanton phenomology; however,
more generally they are consistent with suitable regions of (nearly)
locked E \& B fields.

\begin{figure}
\epsfig{figure=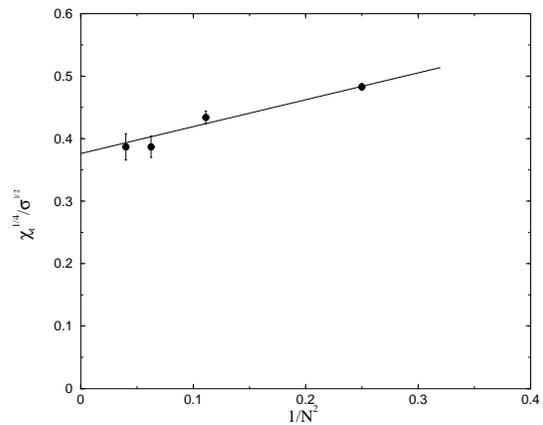, clip=, angle=270, width=7cm} 
\vskip -5mm
\caption{The continuum topological susceptibility in units 
of the string tension plotted against $1/N^2$. A linear
extrapolation to $N=\infty$ is shown (Ref.~\cite{Teper_01}).}
\label{fig:fig_khiN}
\end{figure}

\subsection{Large $N_c$}

Why is there a concern of reconciling the large $N_c$ limit of QCD
with studies of the QCD vacuum? The large $N_c$ description of QCD is
quite successful phenomenologically. It forms a basis for the valence
quark model and OZI rule, and well describes hadron spectra and matrix
elements.  The Witten-Veneziano prediction for relating the
topological susceptibility and meson masses may well be reached at
$N_c=3$.

How do gauge theories approach the limit? The prediction is that for a
smooth limit, one should keep a constant t'Hooft coupling,
$\lambda=g^2 N$ as $N_c\rightarrow\infty$.  Is the limit realized
quickly?  Fig.~\ref{fig:fig_khiN} shows a test for the quenched gluonic topological
susceptibility~\cite{Teper_01}.  The large $N_c$ limit is apparently
realized quickly (seen more definitely in a 2+1 study~\cite{Teper_00})
and is consistent with the predicted $1/N_c^2$ scaling~\cite{Teper_01}.

Revisiting the issue of local chirality, it was found that the amount
of peaking observed for example in Fig.~\ref{fig:hist_IW_2l_2p5pc}
decreases at a fixed scale set by the string-tension as $N_c$
increases from 3 to 4~\cite{Wenger_01}.  This disappearance of local
chirality is consistent with large (instanton-like) modes
disappearing. In fact, Witten~\cite{Witten_79} predicts a strong exponential
suppression of instanton number density.
Teper~\cite{Teper_80} argued that including combinatorial factors for
the measure could compensate for the exponential suppression leading
to a sharp cutoff of the size distribution for small modes, but non-zero
above the cutoff. However, given this limited step
in increasing $N_c$ the decrease in local chirality is consistent with the large $N_c$ predictions. More work is needed to sort out contributions to
the size distribution from dislocations. In any case, the large $N_c$
limit is an interesting place to gain additional understanding of the
QCD vacuum. Chiral fermions have been essential in these studies.

\subsection{Eigenmode Dominance}

How much are hadron correlators dominated by low eigenmodes?  In
recent work~\cite{DH_00,DG_01} comparisons were made of the truncated
and full spectral decomposition of various correlation functions using
the (APE smeared) overlap-Dirac operator.  The lowest 20 modes
(including zero modes) were computed. These new results with a chiral
fermion action draw some of the same conclusions as older
works~\cite{Kilcup_97,Negele_93}. Namely, for small quark masses the
non-singlet pseudoscalar correlator (pion) is well approximated by the
truncated spectral sum.  The non-singlet vector correlator only
saturates at long time distances. These results are consistent with
instanton phenomology.

\begin{figure}
\epsfig{figure=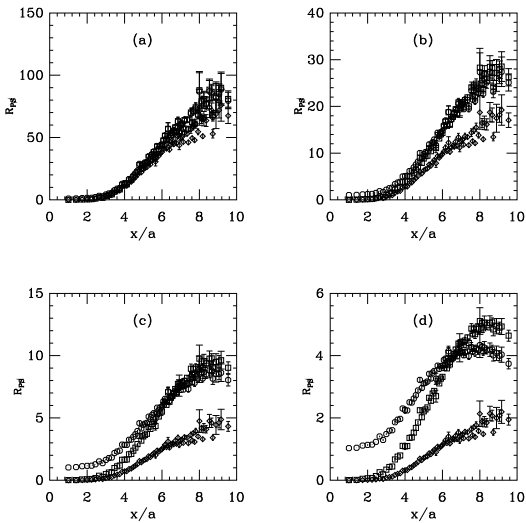, clip=, width=7cm} 
\vskip -8mm
\caption{
Saturation of the point-to-point pseudoscalar correlator by 
 low-lying eigenmodes of the hermitian overlap operator.
 Results for 4 quark masses are shown with $m_{PS}/m_V$
ranging from $0.34$ in (a) to $0.64$ in (d).
 Octagons  show the full hadron correlator.
Squares show the contribution from the lowest 10 modes.
 Diamonds show the contribution from the zero modes,
 which just scales as $1/(am_q)^2$ (Ref.~\cite{DG_01}).
}
\label{fig:pseudo}
\vskip -9mm
\end{figure}

The QCD sum rule approach parameterizes short distance correlators via
the operator product expansion and long distance by condensates.
There are large non-pertubative physics in non-singlet pseudoscalar
and scalar channels. With the overlap-Dirac operator
above~\cite{DG_01}, the truncated spectral sum for the point-point
propagators shows the appropriate attractive channel for the
pseudoscalar and repulsive channel for the vector predicted by
instanton phenomology and consistent with the Wilson
case~\cite{Negele_93}.
\begin{eqnarray}
&&\!\!\!\!\!\!\!\!\!\!\!\!R_i(x) = \Pi_i(x)/\Pi^0_i(x),\quad 
\Pi_i(x)=\Tr\langle J_i^a(x) J_i^a(0)\rangle,\cr
&&\!\!\!\!\!\!\!\!\!\!\!\!J_i^a(x)=\psibar(x)\tau^a\Gamma(i)\psi(x)
\end{eqnarray}
Fig.~\ref{fig:pseudo} shows the ratio $R_i(x)$ for the full and
truncated pseudoscalar case with the free propagator for different
quark masses. Saturation requires few modes for the lightest masses.
One caveat is the use of APE smearing for the gauge links which could
adversely affect the short distance part especially for heavier masses.

\subsection{Screening Correlators}

In the high temperature phase of QCD, one expects restoration of
chiral symmetry exhibited in the limit $m_q a \rightarrow 0$ by the
equivalence of (non-singlet) screening correlators
\begin{equation}
C_S(z) = -C_{PS}(z),\quad C_V(z)=C_{AV}(z)
\label{eq:screening_corr}
\end{equation}
summed over the dimensions orthogonal to the $z$ direction.  A
quenched calculation~\cite{Gavai_01} with the overlap-Dirac operator
at the $N_t=4$ transition shows good agreement in the vector (V) and
axial-vector (AV) channels whereas previous $N_f=0$ \& 2 calculations
did not show agreement in the scalar (S) and pseudoscalar (PS)
channels. The zero mode contributions tend to spoil
Eq.~\ref{eq:screening_corr} but are a finite volume artifact. One can
subtract them out directly or cancel them by comparing
$(C_{S}(Z)-C_{PS}(z))/2$ to $C_{PS}(z)$ without zeromodes. The
agreement without the zero modes tends to support the conclusion that
chiral symmetry is restored and parity doubling is seen. However, the
lack of a diverging chiral condensate seen in Ref.~\cite{KN_01}
probably from a small physical spatial volume indicates that
increasing the size could break the parity doubling once again. More
work to resolve this issue is needed.

\subsection{Localization}

With chiral fermions, one can test the issue of localization of fermionic
eigenmodes and can possibly compare to localization of gluonic quantities.
The latter is however strongly affected by dislocations in the gauge fields
forcing the need for cooling or smearing. This unfortunately can affect
part of what one was hoping to test. However, fermionic 
studies~\cite{DH_00,DG_01,Gavai_01,EHN_99,Regensburg_01} provide
evidence for localization.

\section{Chiral Fermions}

While low eigenmodes of chiral fermions have been discussed, the
implementation of the chiral fermion operators is strongly affected by
low lying modes. For recent reviews of the overlap and
domain wall (DWF) constructions see
Refs.~\cite{Vranas_00,Hernandez_01}. In particular, the
five-dimensional domain wall operator is equivalent to the
four-dimensional overlap after taking the extent of the fifth
dimensional to infinity.  The overlap operator has the form
$D_{overlap}(0) = (1+\gamma_5\varepsilon(H(-M)))/2$ where $H(-M)$
could be the super-critical hermitian Wilson-Dirac operator with mass
$M$.  The numerical implementation of $\varepsilon(H(-M))$ whether by
an extra dimension via DWF or some rational approximation is adversely
affected by small eigenvalues of $H(-M)$. The deviation of
$\varepsilon(H(-M))$ from $\pm 1$ induces chiral symmetry
breaking~\cite{EH_01,CPPACS_01}.  In both the overlap and DWF case one
can project out some number of low lying eigenmodes and correct the
$\varepsilon(H(-M))$ to $\pm 1$~\cite{EH_00}.

The issue of whether there is a large number or even a finite density
of small (zero) eigenvalues of $H(-M)$ is a technical problem related
to Wilson-like gauge actions since it can be shown that given bounds
on a plaquette from unity there is also a corresponding bound on
$\lambda_{min}(H(-M))$~\cite{neubound}.  There are classes of
configurations not satisfying these bounds that induce exponentially
localized zero-modes of $H(-M)$ which occur with non-zero density in a
Wilson-like gauge action, and that have been observed to rapidly
decrease with coupling~\cite{rho0}.
The particularly troublesome modes are from dislocations of the gauge
fields.
The density of these modes is greatly affected by
changing the gauge action and lowering plaquette 
fluctuations~\cite{Orginos_01}. 
In practice, the effects on $\varepsilon(H(-M))$ can be fixed by
projection or using a weaker coupling or both making the question of a
finite density somewhat moot.  However, for a dynamical calculation in
the chiral limit the zero level crossings of $H(-M)$ should be
repelled from crossing at $M$ of the simulation, hence no ambiguity in
the construction of a chiral fermion.

At very strong coupling a new phenomena emerges where the density
of the dislocations becomes so large that the single flavor region
in $M$ merges with the doubler region ($M>2$). Strong coupling and
mean-field calculations differ over this mixing~\cite{strong}, and
the approach to the continuum limit.

\section{Conclusions}

Low lying fermionic modes provide a powerful probe of the vacuum.
There are many studies using fermionic modes in quenched
theories. Obviously more studies are needed with dynamical fermions.

This work was supported by DOE contract DE-AC05-84ER40150 under which the
Southeastern Universities Research Association (SURA) operates the
Thomas Jefferson National Accelerator Facility (TJNAF).

\end{document}